\documentclass[aps,pre,twocolumn,groupedaddress,showpacs]{revtex4}
\usepackage[dvips]{graphicx}

\begin{document}
\title{The role of lubricant molecular shape in microscopic friction}
\author{Oleg M.\ Braun} \email[]{obraun@iop.kiev.ua}
\affiliation{Institute of Physics, National Academy of Sciences of Ukraine, 03028 Kiev, Ukraine}
\author{Nicola Manini}
\affiliation{Dipartimento di Fisica, Universit\`a di Milano,
Via Celoria 16, 20133 Milano, Italy}
\author{Erio Tosatti}
\affiliation{International School for Advanced Studies (SISSA), Via Beirut 2-4, I-34014 Trieste, Italy}
\affiliation{CNR-INFM Democritos National Simulation Center, Via Beirut 2-4, I-34014 Trieste, Italy}
\affiliation{International Centre for Theoretical Physics (ICTP), P.O. Box 586, I-34014 Trieste, Italy}

\begin{abstract}
With the help of a simple two-dimensional model we simulate the
tribological properties of a thin lubricant film consisting of linear
(chain) molecules in the ordinary soft-lubricant regime.
We find that friction generally increases with chain
length, in agreement with their larger bulk viscosity.
When comparing the tribological properties of molecules which
stick bodily to the substrates with others carrying a single sticking
termination, we find that the latter generally yield a larger friction than
the former.
\end{abstract}

\date{Oct. 2, 2008}
\maketitle

\section{Introduction}
\label{intro}

Understanding sliding friction between substrates 
separated by a thin lubricant film is technologically crucial
but also rich physically~\cite{P0,BN2006}.
Without lubricant, it is known that the lowest friction can be achieved for a
contact of two hard crystalline solids, when the atomically flat surfaces are
incommensurate (e.g., see \cite{BN2006} and references therein).
When the surfaces are separated by a lubricant film, the lowest friction
is achieved if the film is solid and crystalline but incommensurate with the substrate surface.
Such a situation appears in the so-called ``hard'' lubricant system
\cite{BN2006}, where the intermolecular interaction strength within the lubricant,
$V_{ll}$, is stronger than the lubricant-substrate interaction, $V_{sl}$.
In that case, the shape of lubricant molecules is not too relevant, because
friction is determined uniquely by the substrate-lubricant interface structure.

Conventional lubricants, however, belong typically to the opposite ``soft''
lubricant type, with $V_{ll} < V_{sl}$.
Here each substrate is covered by a glued lubricant monolayer,
protecting it from wear.
Sliding occurs deeper inside the lubricant film, where
below the film melting point a typical
melting-freezing mechanism operates: the lubricant locally melts at slip (i.e.,
during sliding) and re-solidifies at stick \cite{RT1991}.
For a thick lubricant film above its melting point, e.g.,
and for thicknesses larger than say ten molecular diameters,
friction is proportional to the bulk viscosity of the lubricant.
Therefore, thick fluid lubricants consisting of simple spherical molecules or
individual atoms, characterized by a low bulk viscosity, would provide the
best frictional properties.
Indeed, experiments show that the bulk viscosity for linear molecules
grows with the molecule length as was shown in Ref.~\cite{SSP2003}.
One could then expect that the friction coefficient of chain lubricant
molecules should grow with their molecular length.

However, it is well known that long molecules often make better lubricants.
The reason for this is the squeezout effect: at high load, when the regime
of boundary lubrication is approached and the film is locally extremely thin,
at stick the lubricant may be squeezed out in correspondence of the substrate asperities.
Direct contact of unlubricated asperities leads to a high static friction
force $f_s$ and generally to substrate wear.
Sivebaek {\it et al.}~\cite{SSP2003} showed that with the increase of the
alkane chain length, transitions from $N_l$ to $N_l -1$ lubricant layers
occurs at higher pressure.
Longer lubricant molecules are harder to squeeze out, and this is why they
provide better lubrication properties.
Following that line, Ref.~\cite{SSP2003} studied the ``totally-glued'' case,
where all atoms of the lubricant molecule interact strongly with the substrates.
Longer molecules adhere more to the substrates, making it harder for the squeeze-out of
the last lubricant layer to nucleate.
In agreement with that, longer alkanes perform better as boundary
lubricants than shorter ones, because they prevent more effectively the
appearance of cold-welded junctions and the resulting wear.

Tribological properties of lubricants constructed of linear (chain)
molecules of length $L$, again for the totally-glued system, were also
studied with the help of molecular dynamics (MD) simulation
by He and Robbins \cite{HR2001s}.
They found that at the small coverage $\theta=M/N_s=1/8$ (the number of
lubricant molecules $M$ divided by the number of substrate surface atoms
$N_s$) the dependence of the static friction on chain length
(for $L=1,3,6$) is small, while for a higher coverage, $M/N_s=1/2$, chains
with
$L=3$ and 6 show nearly the same behavior, while ``monoatomic molecules''
($L=1$) display a dramatically reduced friction (about one fourth).

The aim of the present work is to study the role of molecular length in
detail for a wide range of model parameters.
Moreover, we wish to compare a totally-glued lubricant (with all
monomers of the lubricant chain sticking to the substrates) to
a ``head-glued'', with a single (head) atom of the molecule
strongly attached to the substrate, while the other atoms (molecular tail)
interact more weakly with the substrates.
Our aim is to clarify the role of the length of the lubricant molecules in
friction, in particular to find what length of lubricant
molecules could provide best tribological characteristics for a thin (few
molecular layers) soft lubricant film.
We will use the following criteria to define good tribological properties:
(i) a better lubricant provides lower values for both the static
friction force $f_s$ and the kinetic friction force $f_k$, and
(ii) a better lubricant is characterized by lower values for the critical
velocity $v_c$ of the transition from stick-slip to smooth sliding (or at
least to provide a less irregular motion in the stick-slip regime).

With the aim of addressing general trends, we choose to explore a
simplified two-dimensional (2D) minimal model which in our view is
sufficient to catch the physics of the problem, with qualitatively correct trends.
In return for the model's simplicity, we will be able to span a large
choice of structures and parameters. On top of that, the 2D geometry
provides an easier visualization of all processes inside the lubricant.
The 2D model does not describe realistically the lubricant squeezout
\cite{persson_tosatti94},
which requires a full three-dimensional (3D) modeling, such as
that employed in MD simulations of Ref.~\cite{SSP2003}. Therefore
we will not address squeezout properties any further.

The paper is organized as follows: Sec.~\ref{model} introduces our
minimal model; Sec.~\ref{technique} provides the mathematical
and numerical details of the model's solution; Sec.~\ref{linear}
contains a detailed analysis of the results, which are finally
discussed in Sec.~\ref{discussion}.

\section{The Model}
\label{model}

\begin{figure}
\includegraphics[width=82mm,clip=]{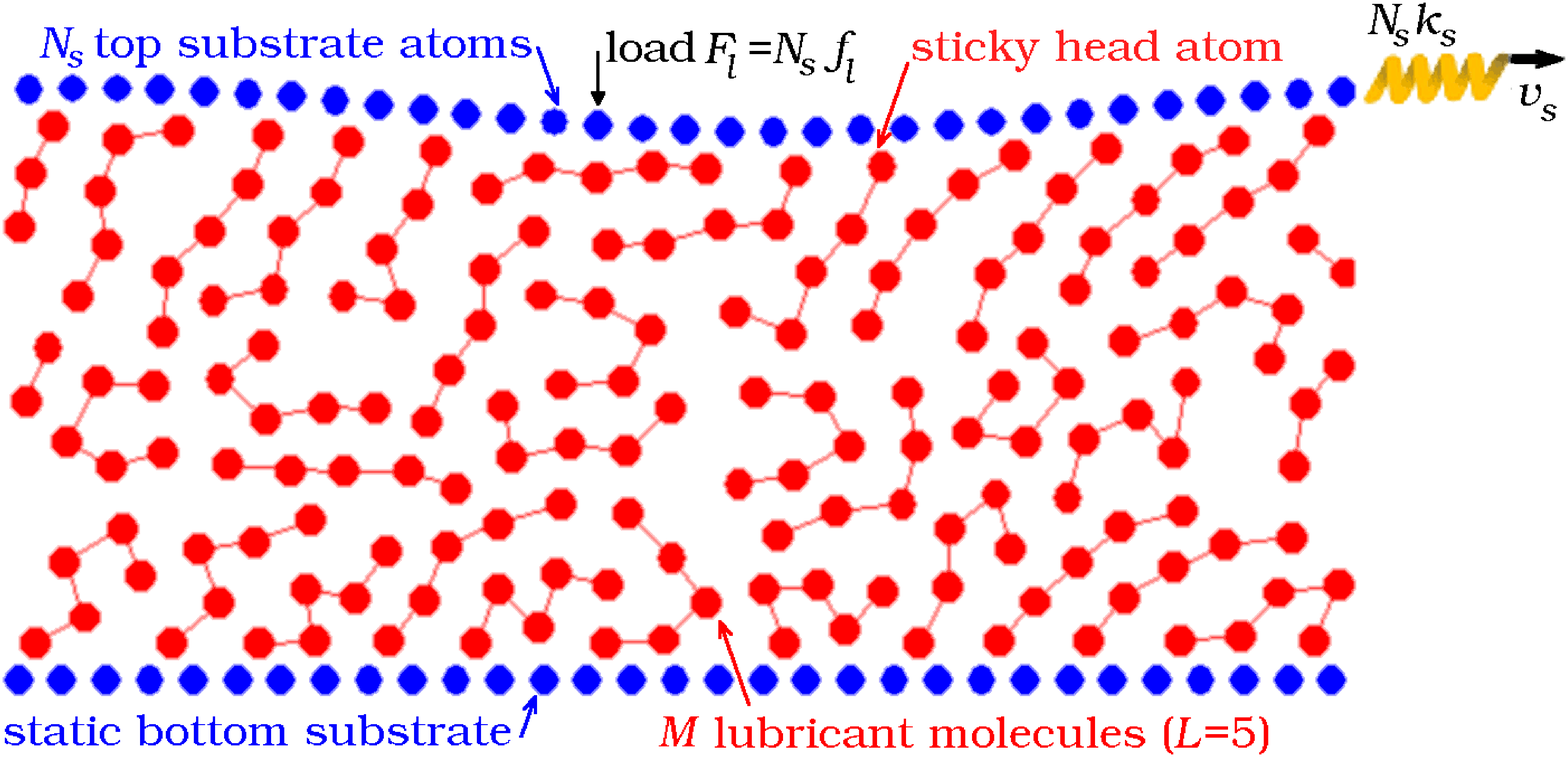}
\caption{\label{model:fig} (Color online)
A schematic cartoon of our simulation model.
This snapshot refers to a configuration where the $N_s=31$-atoms substrates
(blue) squeeze $M=35$ head-glued molecules, each made of $L=5$ atoms (red).}
\end{figure}

We use a 2D model, where point particles, representing individual
atoms or monomer units of larger molecules, can move in two dimensions
$x$ and $z$, where $x$ is the sliding direction and $z$ is perpendicular to the substrates.
The substrates are modeled as two rigid chains of $N_s$ atoms and
lattice spacing $R_s$, so that the total mass of each substrate is $N_s
m_s$ and the system size in the sliding direction is $L_x = N_s R_s$.
We apply periodic boundary condition along the $x$ direction.
The bottom substrate is fixed at $x=z=0$, while the top one is
free to move in both $x$ and $z$ directions.
The top substrate is pressed toward the bottom one by a $(-z)$-directed
constant load force $F_l = N_s f_l$, and is driven in the $x$ direction at
velocity $v_s$ through an attached spring of elastic constant $N_s k_s$, as
sketched in Fig.~\ref{model:fig}.
The spring force $F$, which works against the friction force, is monitored
during simulation (throughout the paper we report the force per substrate atom $f=F/N_s$).
Thus, our model is a 2D variant of a typical experimental setup used in
tribology~\cite{P0,BN2006}.

Between the substrates we insert $N$ lubricant atoms of mass $m_l$.
How eventually these $N$ atoms are to be lumped together to form linear
molecules will be detailed below.
Independently of that, all atoms interact via pairwise 12-6 Lennard-Jones (LJ) potential
\begin{equation}
V_{\rm LJ} (r)=
V_{\alpha \alpha^{\prime}}
\left[ \left( \frac{R_{\alpha \alpha^{\prime}}}{r}\right)^{\!12}
-2 \left( \frac{R_{\alpha \alpha^{\prime}}}{r} \right)^{\!6}\right].
\label{LJ}
\end{equation}
Here $\alpha,\alpha^{\prime} = s$ or~$l$ for the substrate or lubricant
atoms respectively.
Thus, the lubricant-lubricant interaction is characterized by the parameters
$V_{ll}$ and $R_{ll}$, while the lubricant-substrate interaction by
$V_{sl}$ and $R_{sl}$
(direct interaction between the top and bottom substrates is omitted).
Throughout the paper we will use dimensionless units, where $m_s = m_l =1$,
$R_{ll} =1$, and the energy parameters $V_{\alpha \alpha^{\prime}}$
takes values around $V_{\alpha \alpha^{\prime}} \sim 1$.
The relative strength of the energy parameters $V_{ll}$ and $V_{sl}$
determines the low-temperature lubricant behavior \cite{BN2006}.
The lubricant is ``hard'' (i.e., at low temperature it remains solid at
slip) when $V_{ll} \agt V_{sl}$, and ``soft'' (i.e., the film melts at
least locally during slips) for $V_{ll} \ll V_{sl}$.
Because a 2D model cannot reproduce even qualitatively the true phonon
spectrum of a 3D system, and because moreover frictional kinetics is
generally diffusional rather than inertial, we use Langevin equations of
motion with a Gaussian random force corresponding to a given temperature $T$,
and a damping force
\begin{equation}\label{ffriction}
\begin{array}{rcl}
f_{\eta, x} &=&
 -m_l \, \eta (z) \, \dot{x} -m_l \, \eta (Z-z) \, (\dot{x} - \dot{X})
\,, \\
f_{\eta, z} &=&
 -m_l \, \eta (z) \, \dot{z} -m_l \, \eta (Z-z) \, (\dot{z} - \dot{Z}) \,.
\end{array}
\end{equation}
Here $x,z$ are the coordinates of a generic lubricant atom and $X,Z$ is the
center-of-mass coordinate of the top substrate.
The viscous damping coefficient is assumed to decrease exponentially
with the distance from the corresponding substrate,
\begin{equation}\label{etaprofile}
\eta(z) = \eta_0 \left[ 1-\tanh (z/z_d) \right],
\end{equation}
where we typically use $\eta_0 =1$ and $z_d \simeq 1$.
This form is meant to mimic the Joule heat dissipation, which can only
take place through the substrates, and therefore is mostly effective at the
interfaces.
The Gaussian width of the Langevin random force equals $[2 \eta(z) m_l
T]^{1/2}$ \cite{Gardiner}.

We compare the main tribological properties of lubricants  consisting of
molecules of different lengths, from individual (atom-like) monomers to
linear chain molecules composed of $L$ atomic/monomeric units, so that the
number of lubricant molecules is $M=N/L$.
To build chain molecules, we follow the procedure
due to Robbins {\it et al.}~\cite{MR2000,HR2001s}.
The lubricant is described with the Kremer-Grest bead-spring model
\cite{KG1990}, which yields realistic dynamics for polymer melts \cite{T1998}.
While all monomers within a molecule still interact with each other via the LJ
potential, Eq.~(\ref{LJ}), adjacent monomers inside a given molecule (chain)
interact via an additional (``FENE'') potential \cite{KG1990}
\begin{equation}\label{CH}
V_{\rm CH} (r) = \left\{
\begin{array}{ll}
 - \widetilde{V}_{\rm CH} \ln \left[ 1-(\frac r {R_{\rm CH}})^2 \right]
&{\rm for}\ r<R_{\rm CH}
\\
+\infty
&{\rm for}\ r\geq R_{\rm CH},
\end{array}
\right. 
\end{equation}
where $R_{\rm CH} = 1.336 \, R_{ll}$ and $\widetilde{V}_{\rm CH} = 33.75 \,
V_{ll}$ (these parameters are the same as in Ref.~\cite{MR2000}).
This additional potential has the main effect of preventing molecular breaking.
Besides that, it is fairly small at the equilibrium distance $r=R_{ll}$
of $V_{\rm LJ}$, and causes a bond-length contraction to
$R_{\rm mol}\simeq 0.856 R_{ll}$.

\section{Simulation Technique}
\label{technique}

Our frictional simulations use periodic boundary conditions throughout,
a choice which as usual minimizes size effects.
Results are generally dependent on the total number of lubricant atoms $N$.
If $N$ does not match exactly the exact number of atoms required for an
integer number of closely packed layers, then the extra interstitial atoms
or vacancies constitute structural defects.
In real life the presence of these defects is of course the rule
rather than the exception, but in
the large interface size of an actual experiment their importance is more
marginal than in our finite-size simulation.
To reduce errors due to these misfit defects, we use a geometry with a
slightly curved top substrate, whose $z$-coordinate varies along the $x$
direction by
\begin{equation}\label{curvature}
z = Z + \frac 12 R_{sl}  \left[1 - \cos \frac {2 \pi (x-X)}{ L_x}\right],
\end{equation}
as illustrated in Fig.~\ref{model:fig}.

In the simulations to be presented below, each substrate contains $N_s=62$
rigid atoms and the substrate lattice constant is chosen as $R_s =2/3$,
moderately ``incommensurate'' with the lubricant equilibrium interatomic
distance $R_{ll}=1$, a feature not atypical for lubricant/substrate interfaces.

Accordingly, the system size is $L_x = 41.3$.
This fits up to 
$N \alt 50$ atoms (in the atomic lubricant case)
in one monolayer film under a standard applied load $f_l=1$.
The $N = 100$ atoms therefore complete two monolayers (one glued to the bottom
substrate, the other to the top substrate). Atoms in excess of 100 pile up
in between, giving rise to a third, then fourth, etc., layer, consisting
of up to 50 atoms, for the typical load conditions of the simulations.

To guarantee an efficient melting-freezing of the soft lubricant
in correspondence to the stick-slip motion for the ``incommensurate''
lubricant-substrate interface, we require a rather strong inequality
$V_{sl} \gg V_{ll}$. Specifically, we choose $V_{ll}=1/9$ and $V_{sl}=5$.
In detail, we set the lubricant-substrate interaction energies to be
$V_{sl}=5$ in the ``totally-glued'' case, while for the ``head-glued''
system, the large interaction $V_{sl}=5$ applies exclusively to the ``head''
molecular atom, while other atoms in the rest of the molecule (the
``tail'') interact with the substrates with the much weaker $V_{sl}'=1/9$.
With these parameters, the melting temperature of the lubricant is of the
order of $T\approx 0.1$; it is a generally decreasing function of the
lubricant layer thickness, but an increasing function of the chain length
and applied load \cite{BP2003}.

The load force used is typically $f_l=1$ per substrate atom.
Assuming an energy scale $9 V_{ll}\simeq 0.1$~eV and interatomic distances
$R_{ll}\simeq 0.3$~nm (typical of Van der Waals intermolecular
interactions), in a hypothetical 3D geometry such load would correspond to
a pressure of almost 1~GPa.

The top substrate is driven at an average velocity $v_s$ through a
relatively stiff spring of elastic constant $k_s=0.1$.
For the remaining parameters, we take $R_{sl} = R_s$ and $z_d = R_s$.
We run all simulations starting from well-relaxed and annealed initial
configurations, and exclude from averaging any initial transient leading to the
steady dynamical state.
In more detail, we solve the Langevin equations of motion using
a fourth-order Runge-Kutta method
with a time step $\Delta t = \tau_0 /128$,
where
$\tau_0 =2\pi R_{ll} (72\cdot 9 \, V_{ll}/m_l)^{-1/2}$
is an average characteristic period of the LJ interatomic potentials.
The initial configuration is realized by a careful annealing cycle up to
the high temperature $T=2$, followed by a ``relaxation'' run, typically of
$4096\, \tau_0$ or longer, at the appopriate Langevin temperature $T$ and
sliding velocity $v_s$, before measuring the friction force by averaging
over a run typically of the same duration.
The final configuration of every run is saved:
this allows us to restart the run when we need to improve accuracy.
To estimate error bars of various quantities, we split the measuring
trajectory into 30 pieces of equal duration, and then estimate the variance
of the global average through the standard deviation of the averages
carried out over each individual piece.
To verify the results, for some parameters we also made runs for total
integration times and system sizes $2\div 4$ times larger to check that the
results do not change.

\begin{figure}
\includegraphics[width=82mm,clip=]{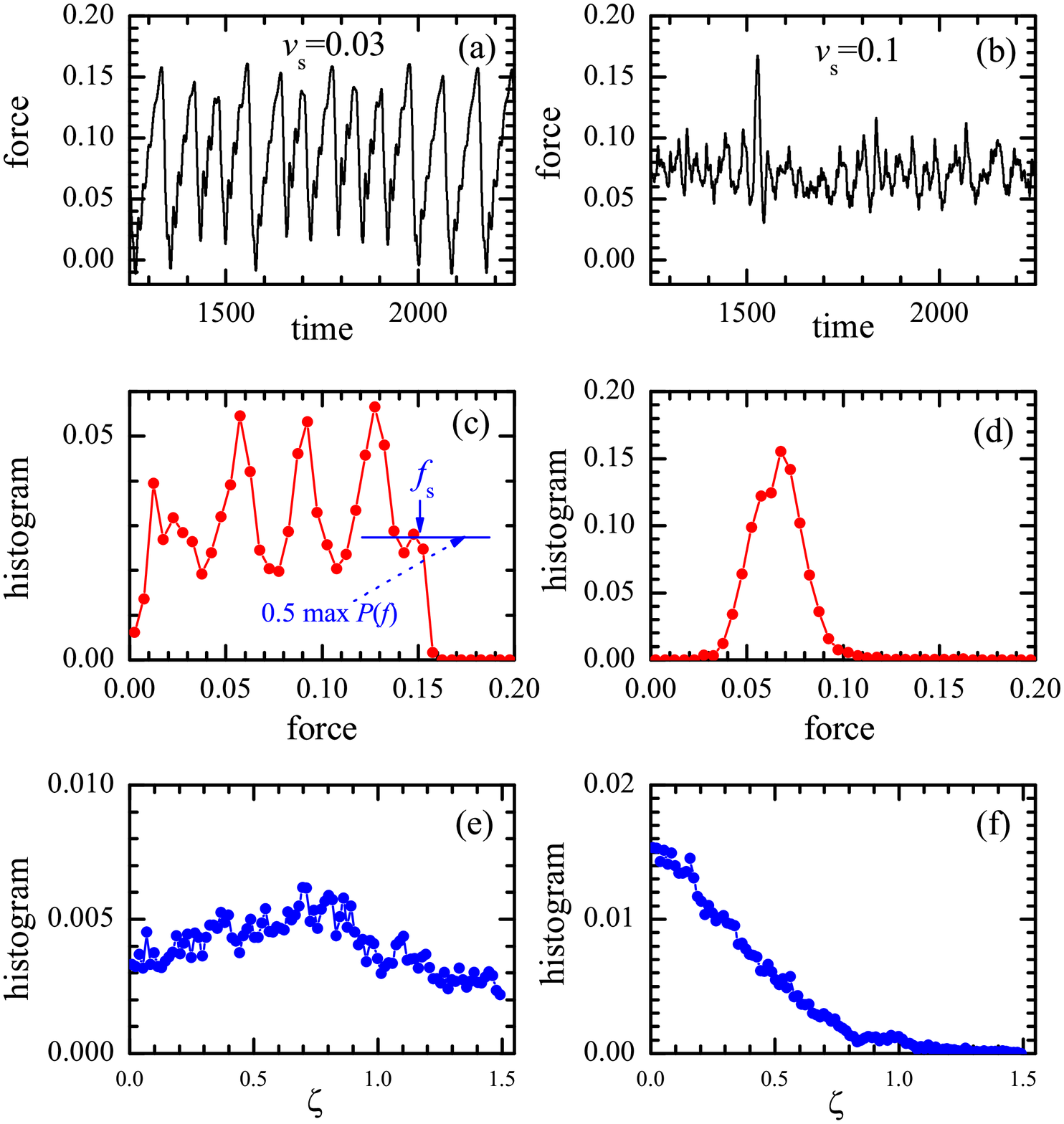}
\caption{\label{L10} (Color online)
Comparison of the stick-slip motion [$v_s=0.03$, panels (a,c,e)]
and the smooth sliding [$v_s=0.1$, panels (b,d,f)]
for $M=40$ totally-glued three-atomic ($L=3$) lubricant molecules at $T=0$.
Panels (a) and (b): typical time dependence of the friction force $f(t)$;
panels (c) and (d): histogram $P(f)$ of the friction force $f(t)$;
panels (e) and (f): depth distribution $P(\zeta)$.
The top substrate is curved, see Fig.~\ref{model:fig}, and the
parameters are the following: $V_{ll}=1/9$ and $R_{ll}=1$, $V_{sl}=5$ and
$R_{sl}=R_s$, $f_l=1$, $k_s=0.1$, $R_s =2/3$, and $N_s =62$.}
\end{figure}

As in typical tribological systems, we observe a transition from
stick-slip to smooth sliding with increasing driving velocity.
An example is shown in Fig.~\ref{L10}.
In the smooth sliding regime, the kinetic friction force can easily be
measured as the average of the spring force $f(t)$.
To evaluate the static friction force $f_s$ in the stick-slip regime, one
can calculate the histogram $P(f)$ of $f(t)$ and
then select $f_s$ as the point where $P(f)$ decays below a given
level, e.g., $P(f_s) = 0.5 \, \max P(f)$ (see Fig.~\ref{L10}c).
A harder problem is that of determining the precise transition from stick-slip
to smooth sliding and its critical velocity $v_c$,
since this transition is often a smooth crossover.
For this purpose, we calculate the distribution (histogram) $P(\zeta)$ of an
auxiliary variable $\zeta$, defined as
$\zeta(t) = \left[ df(t)/dt \right] / \left( k_s v_s \right)$.
In smooth sliding the function $P(\zeta)$ decays monotonically from a
maximum at $\zeta \approx 0$ regime (see Fig.~\ref{L10}f).
In contrast, for stick-slip, $P(\zeta)$ should show a maximum at some $0<
\zeta \alt 1$ (Fig.~\ref{L10}e).  Indeed, during sticks the force grows as
$f(t)\simeq k_s v_s t$ [so that $P(\zeta)=\delta(\zeta -1)$], and it drops
quickly to substantially smaller values during short slip events.
Thus, this method allows us to evaluate $v_c$ fairly accurately.

\section{Results: totally-glued versus head-glued molecules}
\label{linear}

\begin{figure*}[t]
\includegraphics[width=16cm,clip=]{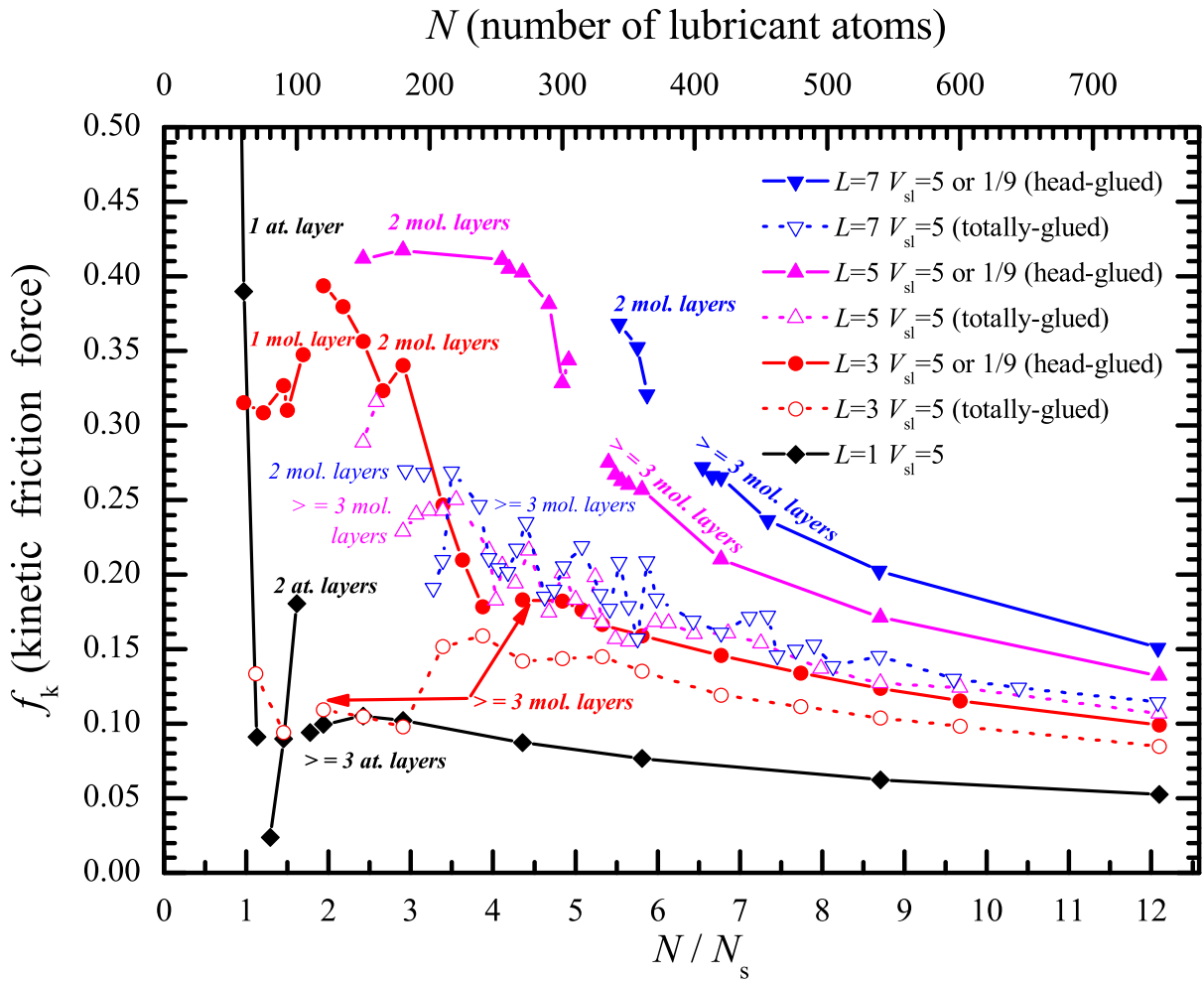}
\caption{\label{L01} (Color online)
Smooth sliding ($T=0$, $v_s=1$)
time-averaged kinetic friction force $f_k$ as a function of
dimensionless coverage $N/N_s$ (bottom scale), as well as total number
of atoms $N=ML$ (top scale) for linear lubricant molecules of different
lengths $L=1$ (black diamonds), 3 (red circles), 5 (magenta up triangles),
and 7 (blue down triangles), for either the totally- (open symbols and
dotted lines) or head-glued (solid symbols and curves) cases.
The top substrate is curved according to Eq.~(\ref{curvature})
(see Fig.~\ref{model:fig}), and the other
parameters are the following: $V_{ll}=1/9$ and $R_{ll}=1$,
$V_{sl}=5$ or 1/9 and $R_{sl}=R_s$, $f_l=1$, $k_s=0.1$, $R_s =2/3$, and
$N_s =62$.}
\end{figure*}

For the chosen set of parameters, the transition from stick-slip to smooth
sliding occurs at a velocity $v_s <1$; therefore, we select $v_s=1$ to
represent the smooth-sliding regime, where we calculate the kinetic
friction force.
Figure~\ref{L01} summarizes the simulation results for $T=0$, reporting the
time-averaged kinetic friction force $f_k$ as a function of the total
number of lubricant atoms, which is basically proportional to the lubricant
layer thickness, for different molecular lengths $L$.

The monoatomic lubricant (diamonds in Fig.~\ref{L01})
forms two complete monolayers for $N=100$
and three monolayers for $N=150$.
For $N \geq 150$, one ordered layer sticks to the bottom substrate and
another ordered monolayer to the top substrate, while the layer(s) in
between are liquid during sliding.
For the monatomic lubricant, the static friction force (not shown)
is larger than the kinetic friction by a factor
of $2$ to $4$; the transition from stick-slip to smooth sliding occurs at
$v_c \agt 0.03$.
Friction generally decreases as the number of lubricant atoms $N$ -- and thus
the lubricant thickness -- grows, but the dependence $f_k (N)$ cannot be
described by the simple laws of viscous-friction flow.
At lower lubricant thickness
$100 < N < 150$, when the middle layer
is incomplete, friction is slightly lower than for three complete
monolayers, see Fig.~\ref{L01}.
When the lubricant amount decreases further, $70 < N \leq 100$,
there remain only two, generally incomplete, lubricant layers,
each glued to the nearest
substrate (e.g., $N=100$ atoms form exactly two complete lubricant
layers, which slide slowly, thanks to incommensurability, over the
substrates to which they are attached).
At even smaller values of $N$ we have one monolayer or
eventually just a monolayer island
in the narrow gap between substrates, and at this point both the
friction and the critical velocity $v_c$ grow rapidly when $N$ decreases.

\begin{figure}
\includegraphics[width=82mm,clip=]{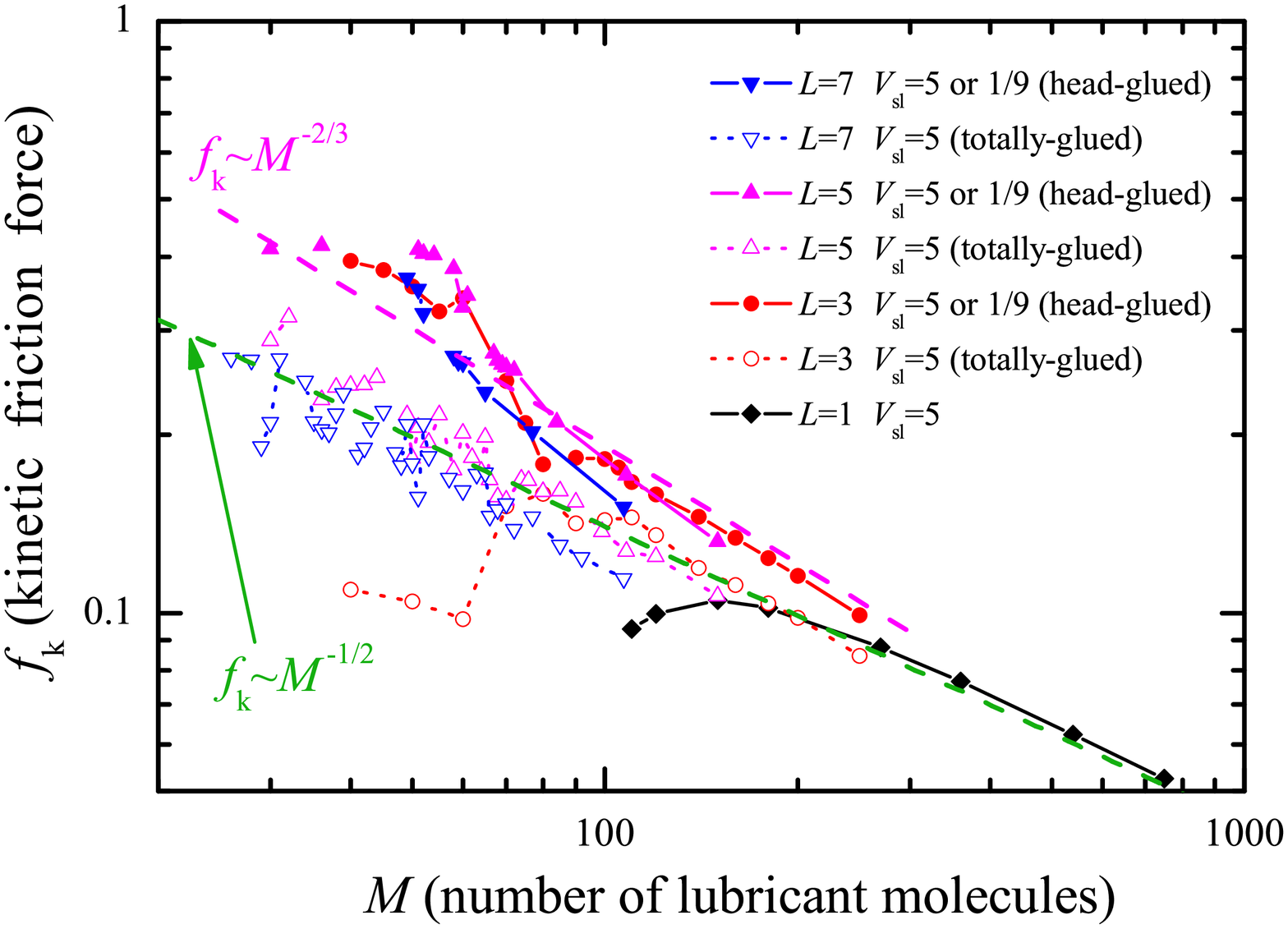}
\caption{\label{L02} (Color online)
Smooth sliding.
Kinetic friction force $f_k$ as a function of the number of lubricant
molecules $M$ (of different lengths $L$) in a log-log scale.
Parameters and notation are as in Fig.~\ref{L01}.
Dashed lines describe power-law fits.}
\end{figure}

For all chain molecules (polymer lubricant) $L>1$,
{\em friction is always larger than for the monomer lubricant.}
The kinetic friction force increases with the chain length $L$,
if the total number of atoms $N$ is kept fixed.
In general, regardless of $L$, friction decreases roughly monotonically
with increasing $N$, the lubricant thickness.
In Fig.~\ref{L02} we re-organize the kinetic friction $f_k$ data of
Fig.~\ref{L01}, but as a function of the number of molecules~$M$ (only
points corresponding to three or more molecular layers are shown).
In the standard viscous flow of a fluid with a viscosity $\widetilde\eta$,
the friction force $f_k$ should depend on the thickness $d=\langle Z
\rangle$ of the liquid layer ($d \propto N \propto M$) as $f_k =
\widetilde\eta v_s R_s^2 /d$.
%
Unlike expectations for a fluid lubricant at high temperature, at $T=0$
the log-log plot of the function $f_k (M)$ suggests that the
simulation data may be (crudely) fit by power laws
$f_k \approx 0.18 \, (M/N_s)^{-1/2}$ for the totally-glued case, and
$f_k \approx 0.26 \, (M/N_s)^{-2/3}$ for the head-glued system.
Accordingly, the function $f_k (M/N_s)$ displays a nontrivial apparent
``universal scaling'' with thickness, at least within the nanoscale
thickness range considered here.
%
The slower decay of friction compared to $d^{-1}$
may be attributed to a non-uniformity
of local heating due to sliding-induced local shear. At fixed driving
velocity, the local shear decreases as thickness increases,
whence the local viscosity involved increases, and so does frictional
dissipation, causing a positive deviation from the
fluid-like $M^{-1}$ behavior.
In other words, the higher average kinetic energy, and therefore higher
effective
temperature, produces a smaller viscosity $\widetilde\eta$
\cite{Seeton2006}, thus a smaller friction for a thinner lubricant layer
than for a thicker one, moderating the power-law decay of $f_k$ in Fig.~\ref{L02}.
For a much thicker lubricant, a $T=0$ simulation would eventually show
lubricant solidification with slip movements associated to local melting,
even at the very high sliding velocity considered here.

\begin{figure}
\includegraphics[width=82mm,clip=]{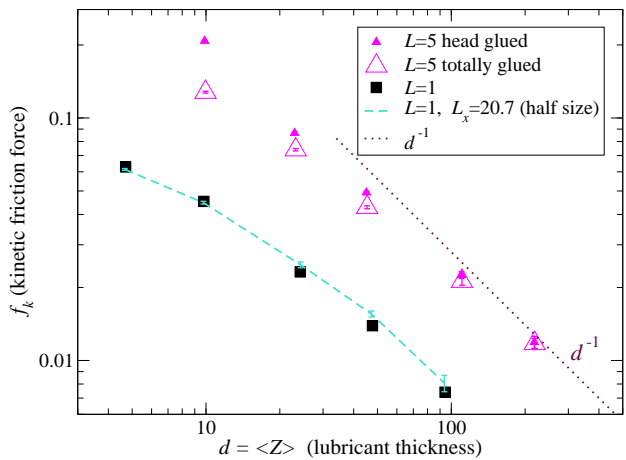}
\caption{\label{thicknessscaling} (Color online)
Kinetic friction force $f_k$ as a function of the lubricant thickness
$d=\langle Z\rangle$ (roughly proportional to $N$) in log-log scale.
The parameters and notation are as in Fig.~\ref{L01}, except for $T=0.2$.
The dotted line sketches the $f_k\propto d^{-1}$ thick-lubricant regime
characterized by a constant viscosity $\widetilde\eta$.
The dashed line with error bars illustrates  the effect of the finite
simulation size on $f_k$ in the $L=1$ case.}
\end{figure}

Figure~\ref{thicknessscaling} reports similar calculations carried out at a
finite temperature $T=0.2$, compared to which the extra kinetic energy
induced by shearing at sliding velocity $v_s=1$ becomes neglible when the
lubricant thickness exceeds about $100\,R_{ll}$.
When the lubricant thickness exceeds such values we do observe a crossover
to the expected scaling $f_k \propto (M/N_s)^{-1}\propto d^{-1}$, as soon
as the shearing kinetic energy becomes smaller than the thermal
fluctuations imposed by the finite $T=0.2$.
Note that small finite-size effects induced by the periodic simulation box
tend to delay slightly the onset of the $f_k \propto d^{-1}$
constant-viscosity regime.

Analyzing the simulation movies for the {\it totally-glued} lubricant with
$L>1$ and large enough $N$ we observe two complete layers sticking to the
corresponding substrates and a liquid lubricant between these layers.
However, not all boundary molecules are glued completely,
some of them are glued partially, i.e.\
only a part of a molecule sticks to the substrate
while another part is detached and points into the lubricant bulk.
It is precisely these ``free tails'' that make the friction higher
as compared with the monoatomic case.
At lower lubricant thickness
($M \alt 50$ for the $L=3$ case,
$M \alt 35$ for the $L=5$ case, and
$M \alt 25$ for the $L=7$ system)
the middle layer is incomplete: its molecules group together at the
narrowest intersubstrate gap and just ahead of it (jamming/plowing effect).
With a further decrease of lubricant thickness, sliding takes place
at a single interface only, namely between the lubricant layers which are
attached to bottom and top substrates.
In this case, we observe a slow ``caterpillar'' motion of glued
molecules relative the substrate, but this effect is weak because of the
strong lubricant-substrate interaction.
At even lower lubricant thickness occasional molecules may be glued to both
surfaces simultaneously, and in such cases the molecular end sticking with
fewer atoms slides along the corresponding substrate.

We come now to analyze the {\it head-glued\/} lubricant, comparing its
behavior to the totally-glued one described above, all other parameters being the same.
Firstly, the head-glued system
(solid symbols in Figs.~\ref{L01} and~\ref{L02})
display a systematically higher friction
than the totally-glued lubricant (open symbols).
As in the totally glued case, the surfaces are covered by lubricant layers.
Again, the layers are orientationally ordered: in this case
the molecular heads are attached to the corresponding surface,
while the tails are directed outside the surface,
and inclined in the sliding direction, like combed hair.
Also, the sliding interface is more ``irregular'',
than in the totally-glued system.
Therefore, one may expect a smoother frictional
motion even in the stick-slip regime at low driving velocities.
It is important to note that contrary to the fully glued case, in the
head-glued system we find essentially no jamming effects even at low
lubricant thickness.

\begin{figure}
\includegraphics[width=82mm,clip=]{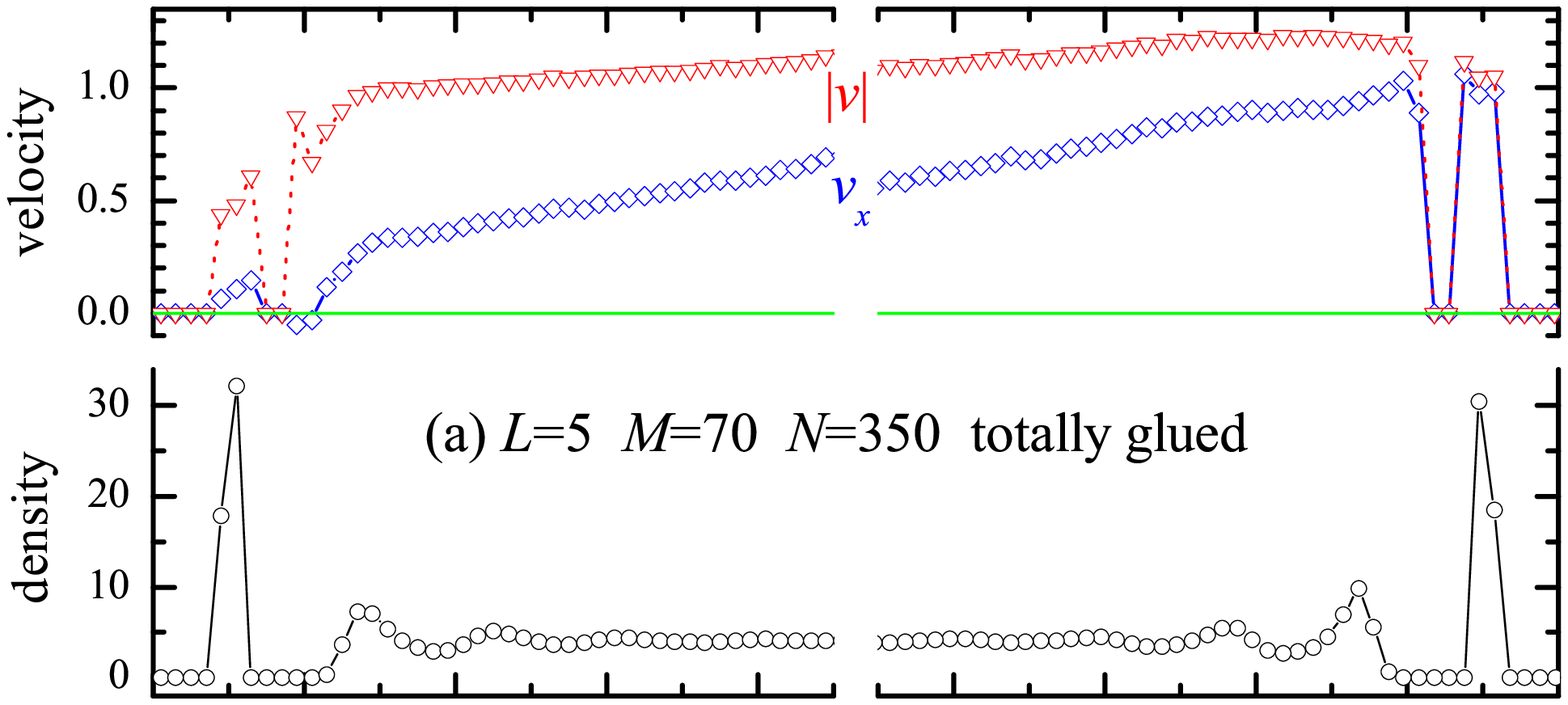}
\includegraphics[width=82mm,clip=]{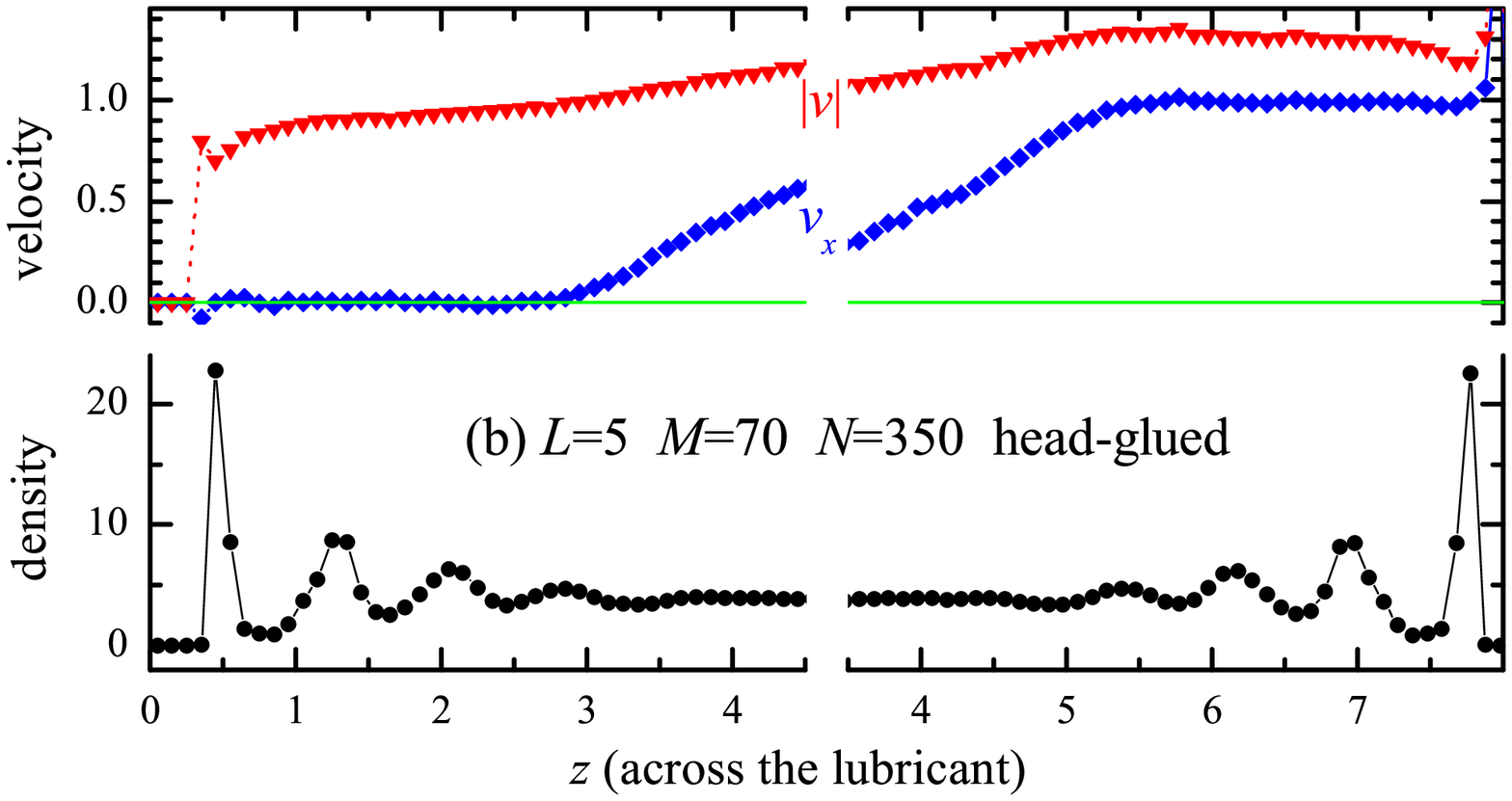}
\caption{\label{L04} (Color online)
$z$-distribution of the average density (circles) and atomic
velocities for (a) the totally-glued molecules (open symbols) and (b) the
head-glued (solid symbols) system with $L=5$ and $M=70$ ($N=350$; other
parameters as in Fig.~\ref{L01}).
Triangles stand for the total velocity $|v|$, diamonds for the
$x$-component of the velocity.
In the right-side panels $z$ is measured with respect to the (fluctuating)
top layer position $Z$, and then translated by its time average $\langle
Z\rangle$. The apparent discontinuity at the center is therefore not real.}
\end{figure}

Figure~\ref{L04} depicts the different atomic density $\rho (z)$ and
velocity distributions $v_x(z)$ and $|v(z)|$ across the lubricant for
the totally- and head-glued molecular lubricants.
In both cases, one layer sticks to the substrate, but the thickness of this
layer is $\simeq R_{ll}$ in the totally-glued system and much thicker,
$\alt L\,R_{\rm mol}$, similar to the molecular gyration radius, in the head-glued case.
Accordingly, while the sliding profile $v_x(z)$ is fairly linear in the
totally-glued case, in the head-glued case the shearing dissipation
concentrates mostly in a bulk lubricant region remote from the substrates.

\begin{figure}
\includegraphics[width=82mm,clip=]{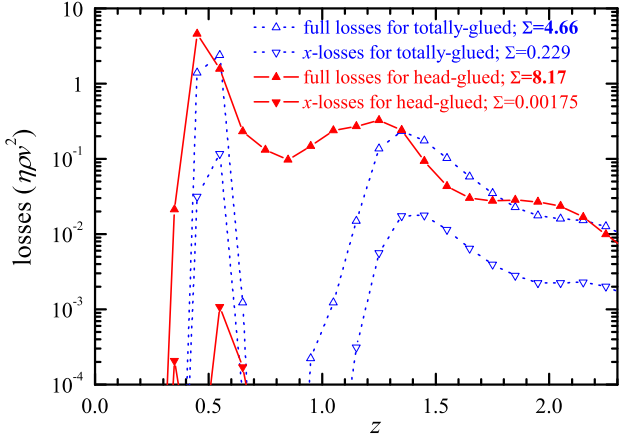}
\caption{\label{L05} (Color online)
$z$-distribution of the total energy losses $\eta(z)\, \rho(z)\,
v^2(z)$ (up triangles) and the $x$-losses $\eta(z)\, \rho(z)\, v_x^2(z)$
(down triangles) for the totally-glued (blue open triangles) and head-glued
(red solid triangles) lubricants for the model parameters as in Fig.~\ref{L04}.
The $z$-integrated energy loss $\Sigma$ is indicated for each curve.}
\end{figure}

To elucidate the different tribological properties of the two kinds of
molecular lubricants, in Fig.~\ref{L05} we analyze the distribution of
energy losses (dissipated power) due to sliding, with the method described
in Ref.~\cite{BN2006}.
The losses occur mainly within the first (glued) lubricant layer, as a
consequence of the model Eqs.~(\ref{ffriction}) and (\ref{etaprofile}),
mimicking the capability of different lubricant layers to transfer energy
into phononic/electronic degrees of freedom of the substrates.
Importantly, we find that the main losses come from atomic vibrations in
the transverse ($z$) direction, by far dominant over those coming from the
motion along the driving ($x$) direction.
This observation explains why, even though head-glued molecules displace
sliding away from the substrates surfaces where most dissipation takes
place, the head-glued system leads to a higher friction: transverse
vibrations of larger amplitude penetrate better near the substrates
(where most dissipation occurs) through head-glued molecules
than through totally-glued ones.

\begin{figure}
\includegraphics[width=82mm,clip=]{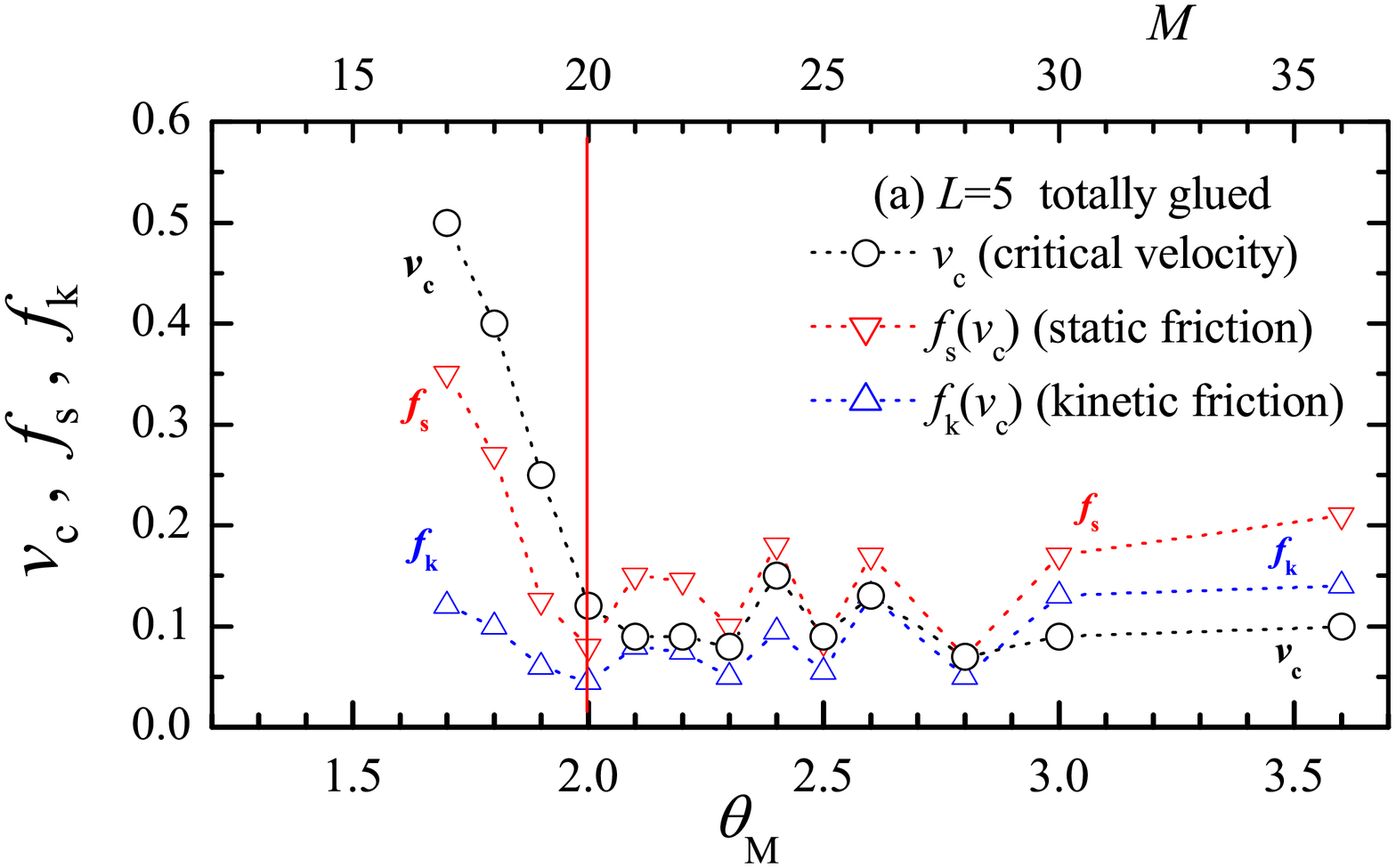}
\includegraphics[width=82mm,clip=]{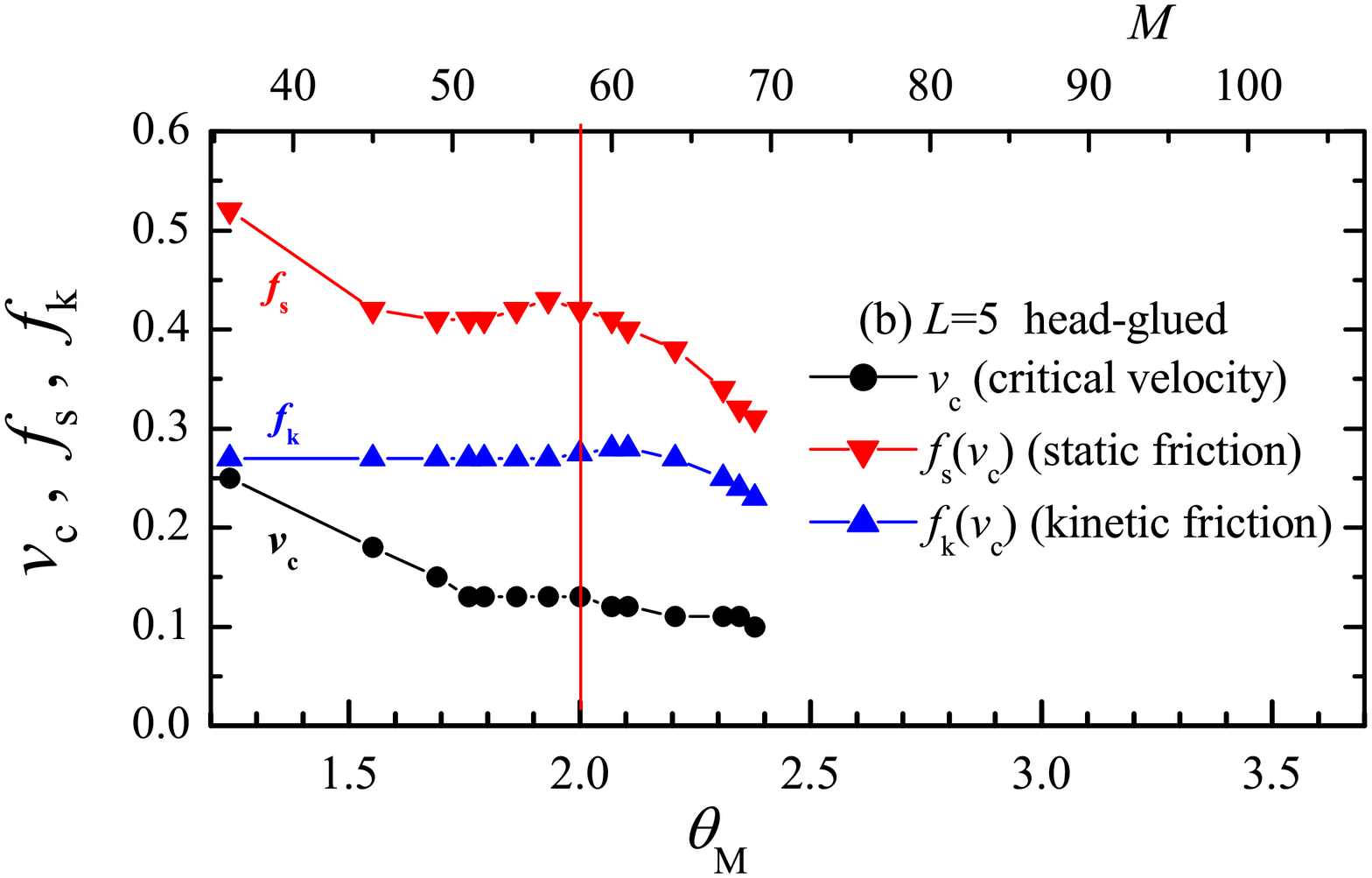}
\caption{\label{L06} (Color online)
Dependence of the critical transition velocity $v_c$ (circles) between
stick-slip and smooth sliding, of the static
friction force (red down triangles) and kinetic
friction force (blue up triangles) at the crossover velocity $v_s=v_c$ upon
the dimensionless coverage $\theta_M$ (bottom scale), or the number of
lubricant molecules $M$ (top scale) for the $L=5$ chain molecules, for
(a) totally-glued and (b) head-glued molecules.
Simulation parameters same as in Fig.~\ref{L01}.}
\end{figure}

Our results suggest that a totally-glued lubricant is more effective
than a head-glued one, because it provides lower kinetic friction.
According to Sivebaek {\it et al.}\ \cite{SSP2003}, totally-glued molecules
with their stronger sticking to the substrate, 
should also be more stable against squeezout of the contact area at high load.
Unfortunately, as was said in the beginning, in our simplified 2D model
we cannot study quantitatively the squeezout process.
Squeezout involves nucleation of a small  $(N-1)$ negative island or ``crater''
and its subsequent expansion, a process which also involves elasticity
of the substrates, and which is very characteristically three-dimensional
\cite{persson_tosatti94}.
We can, however, analyze in detail the dependence of frictional forces on
the number of lubricant molecules $M$.
Results for the $L=5$ lubricant molecules are presented in Fig.~\ref{L06}.
It is useful to define the maximum number of molecules $M_2$ which can
arrange themselves exactly to form the two interface molecular layers (one attached to the
bottom substrate and the other to the top substrate), so that sliding takes
place in the remaining lubricant just in between these two layers.
For the totally-glued lubricant
$M_2\approx 2 L_x/[R_{ll}+(L-1)R_{\rm mol}]$,
and for the head-glued lubricant
$M_2 < 2 L_x/R_{ll}$.
$M_2$ depends on $L$ and on the load force, so that it takes an analysis of
the MD 
trajectories to define its precise values.
In the current model (with $L_x=41.3$, $N_s =62$, $L=5$) we find $M_2=20$
for the totally-glued system and $M_2=58$ for the head-glued case.
Based on the definition of $M_2$ we can introduce the dimensionless coverage
$\theta_M = 2 \, M/M_2$, so that $\theta_M =2$ corresponds to precisely two
molecular layers.
Figure~\ref{L06} shows that, as expected, both the friction force and
the critical velocity $v_c$ increase when the coverage decreases
below the two-layer value.
The rise of friction below  $M_2$ = 2 is fairly slow in the head-glued system,
but is very sharp in the totally-glued case.
As is well known \cite{P0}, in real 3D system the lubricant is squeezed
out from the contact areas at the asperities, where about $2-3$ lubricant
boundary layers are left, harder to remove.
When however the squeezing force is pushed further leaving less than two
layers, the head-glued lubricant continues to operate fairly efficiently,
while the totally-glued lubricant quickly loses its tribological properties.
A similar behavior occurs for chains of length $L=3$ and~7.

\begin{figure}
\centerline{
\includegraphics[height=64mm]{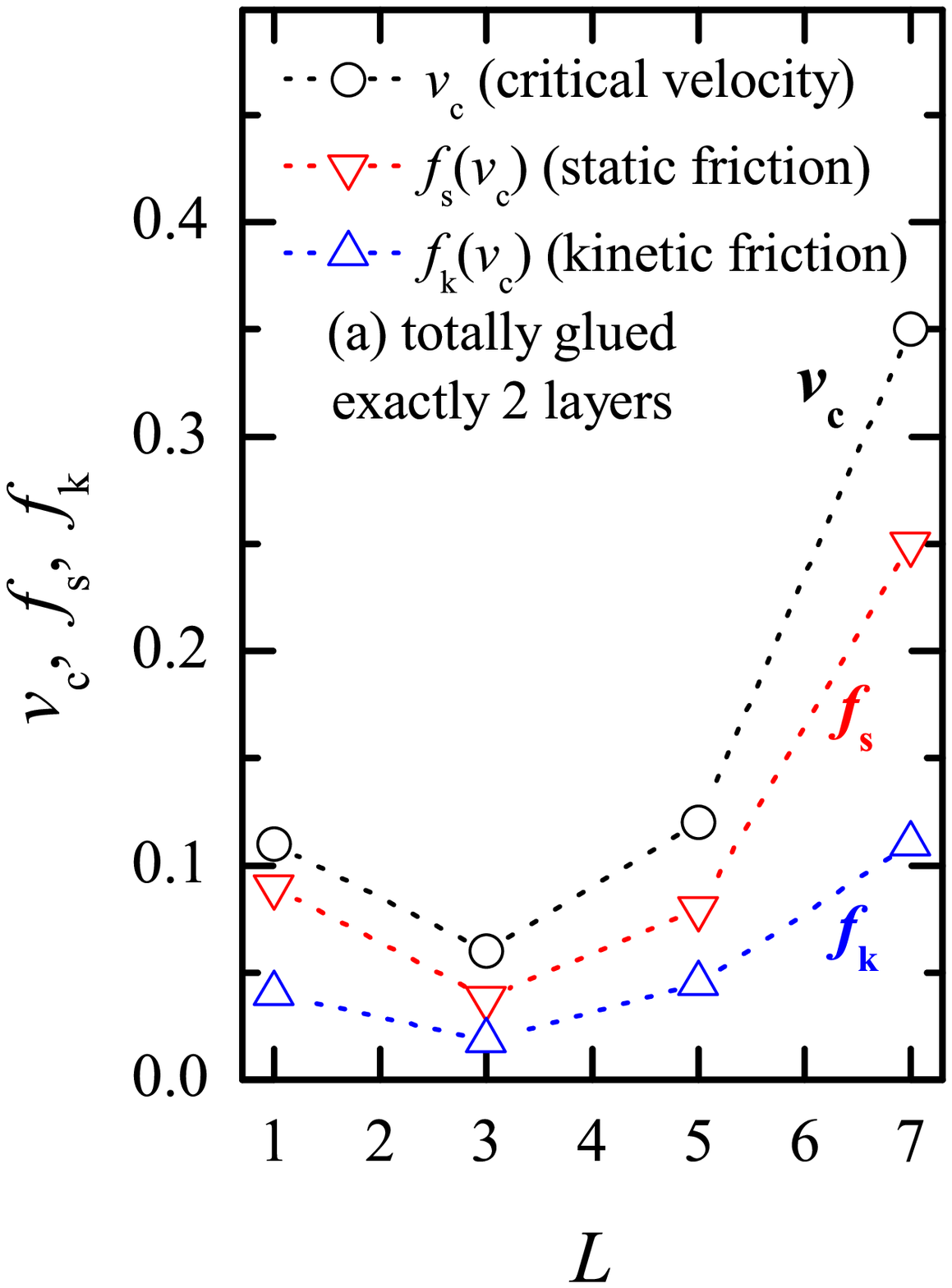}
\hfill
\includegraphics[height=64mm]{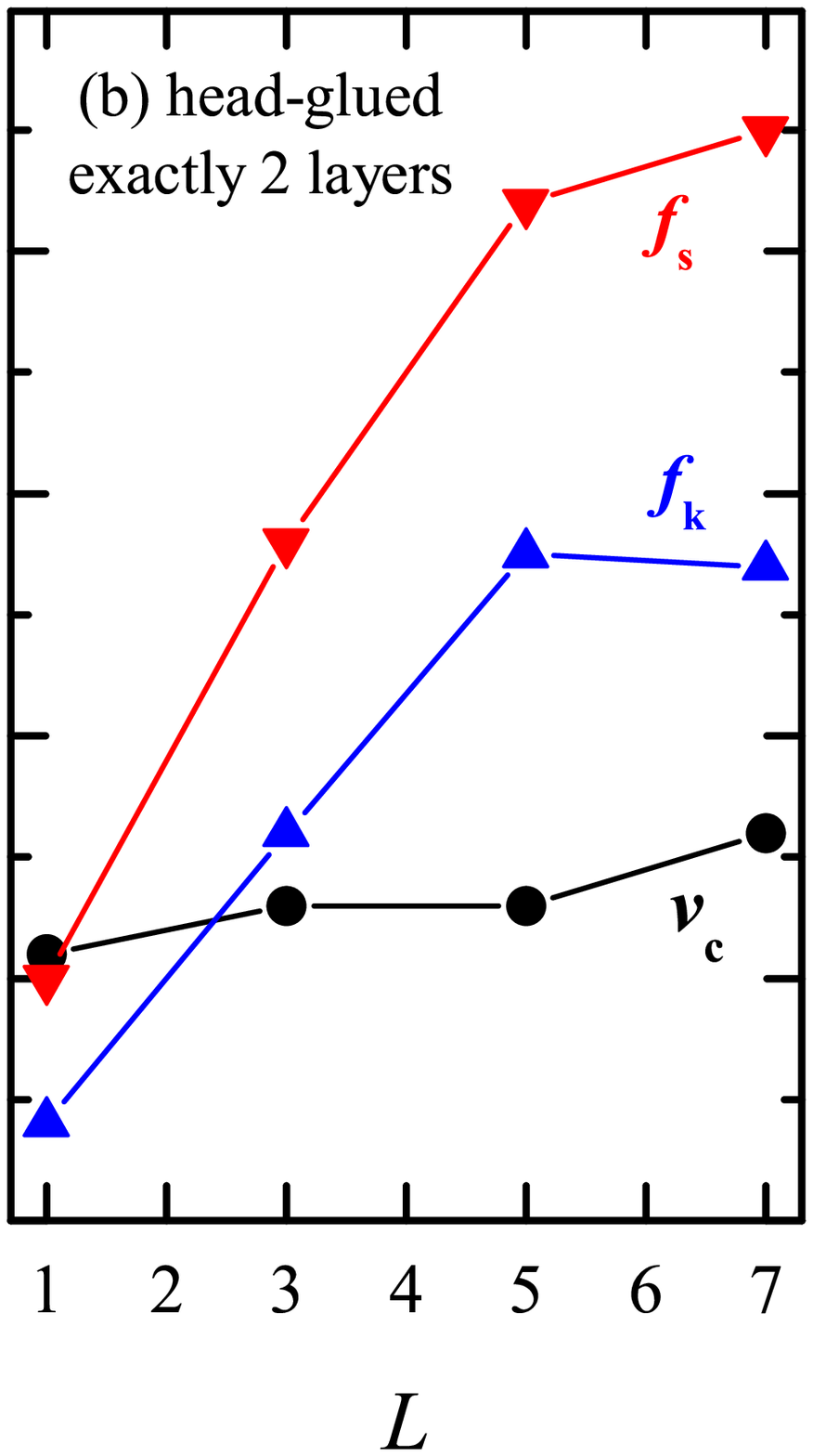}
}
\caption{\label{L07} (Color online)
Dependence of the static (red down triangles) and kinetic (blue up
triangles) friction forces at $v_s \approx v_c$ and of the critical velocity
$v_c$ (circles) on the length $L$ of lubricant chains for the coverage
$\theta_M$ of two molecular layers, in the cases of (a) the totally-glued and
(b) the head-glued system.
Other parameters same as in Fig.~\ref{L01}.}
\end{figure}

The $L$-dependence of friction forces is illustrated in
Fig.~\ref{L07} for a configuration with exactly two molecular layers ($\theta_M=2$).
In detail, the critical velocity $v_c$ marks the crossover illustrated in
Fig.~\ref{L10}, and the frictional forces $f_s$ and $f_k$ are evaluated
immediately below and above $v_c$, in the stick-slip motion and smooth
sliding regimes respectively.
Note that for the head-glued lubricant, the critical velocity depends only
weakly on the molecular length $L$, while in the totally-glued case, $v_c$
increases rapidly with $L$, so that for long molecules, $L > 5$, the
critical velocity of the totally-glued chains is much higher than for the
head-glued ones.
Note also that the static friction $f_s$ for the stick-slip motion at $v_s
\alt v_c$ is approximately twice as large as the kinetic friction $f_k$ in
the smooth-sliding regime at $v_s \agt v_c$.

\begin{figure}
\includegraphics[width=82mm,clip=]{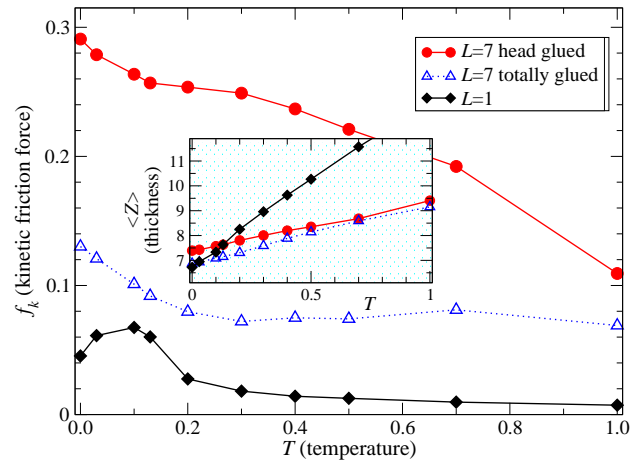}
\caption{\label{Tdependence} (Color online)
Temperature dependence of the kinetic friction force $f_k$ of the
monoatomic lubricant ($L=1$, $N=357$, $M_2 =100$, diamonds) and of head-glued (full
circles) and totally-glued (open
triangles) chain molecules of length $L=7$
($M=51$, $N=357$, $M_2 =14$), at fixed driving velocity $v_s=0.316$.
The inset shows the effective thermal expansion of the mean lubricant
thickness.
Error bars are smaller than the symbol size.
Other simulation parameters same as in Fig.~\ref{L01}.}
\end{figure}

We come now to discuss in some detail how the tribological properties vary
with the system parameters.
Consider first the role of temperature.
Figure~\ref{Tdependence} reports the $T$-dependence of kinetic friction
at a fixed driving velocity $v_s =0.316$, in the smooth sliding regime.
Friction generally decreases when temperature grows, as is to be
expected of the viscosity decrease of a fluid~\cite{Seeton2006}.
General trends for different lubricant molecules are similar:
the head-glued lubricant provides the largest friction, the
monoatomic lubricant the lowest friction.
It is interesting that the temperature dependence of kinetic friction $f_k$
for the totally-glued lubricant is significantly weaker than for the
head-glued and monoatomic lubricants. The same type of dependence was
observed for different driving velocities.
The monoatomic lubricant friction actually increases with $T$,
at low temperature, where the lubricant structure is crystalline.
%
This is expected for an incommensurate lubricant/substrate interface
\cite{BN2006}.
The thermal expansion of the simulated lubricant is rather large,
especially for the monoatomic lubricant, as shown in the inset of Fig.~\ref{Tdependence}.

\begin{figure}
\includegraphics[width=82mm,clip=]{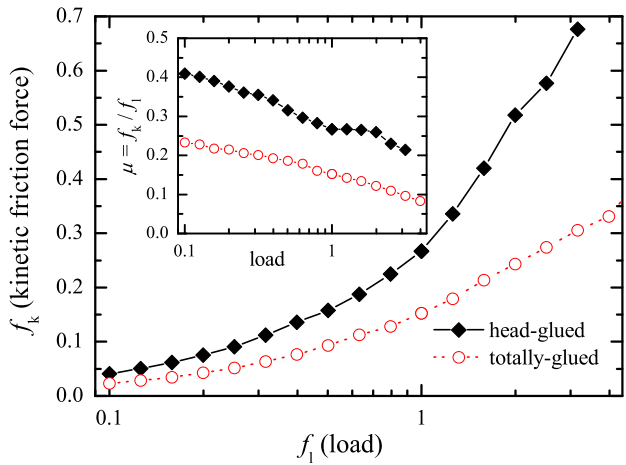}
\caption{\label{L09} (Color online)
Dependence of the kinetic friction force $f_k$ and the friction coefficient
$\mu_k = f_k /f_l$ (inset) on the load force $f_l$ for the totally-glued
(dotted curve and open diamonds) and head-glued (solid curve and diamonds)
lubricant.
Data refer to $M=70$ chains of $L=5$ atoms, other parameters same as in Fig.~\ref{L01}.
Note the semilog scale.}
\end{figure}

Next, Fig.~\ref{L09} shows a typical load dependence of the kinetic friction.
The friction force increases with load, but not quite linearly, i.e., the
friction coefficient $\mu_k = f_k /f_l$ is not
constant (unlike Amontons first law): $\mu_k$ decreases gradually the load
$f_l$ increases.
The reason is most likely associated with the increasing effective shearing temperature
associated with a fixed shearing velocity under increasing lubricant compression,
the lower viscosity in turn reducing friction.
Note that friction of the head-glued lubricant is approximately twice
that of the totally-glued system.

\begin{figure}
\includegraphics[width=82mm,clip=]{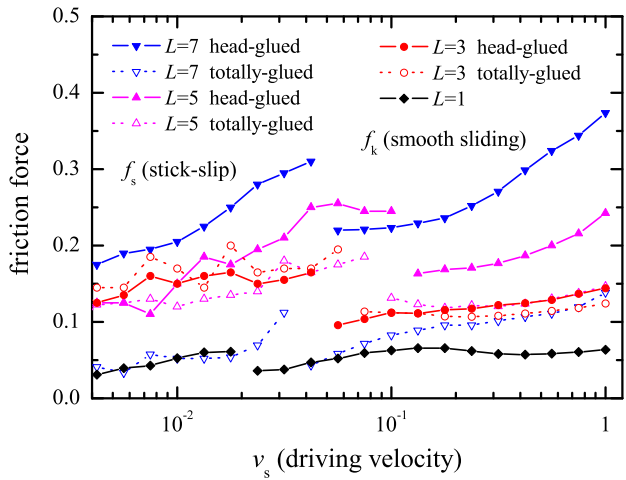}
\caption{\label{L08} (Color online)
Friction forces $f_s$ and $f_k$ as functions of the driving velocity
for linear lubricant molecules of different lengths
$L=1$ ($M=N=360$, solid diamonds),
$L=3$ ($M=120$ and $N=360$, red circles),
$L=5$ ($M=70$ and $N=350$, magenta up triangles), and
$L=7$ ($M=51$ and $N=357$, blue down triangles),
for either totally- (open symbols and dotted curves) or head-glued (solid
symbols and curves) molecules.
The calculation is carried out at temperature $T=0.1$,
other parameters same as in Fig.~\ref{L01}.}
\end{figure}

Figure~\ref{L08} reports the dependence of the friction forces $f_s$
and $f_k$ on the driving velocity $v_s$ at a nonzero temperature $T=0.1$.
The friction forces generally rise with velocity, but very gently so.
By comparison with Fig.~\ref{L07}, we note that the transition from
stick-slip to smooth sliding at a nonzero temperature and for more than two
layers occurs earlier than at $T=0$ and for exactly two layers.
The transition shows a nonmonotonic dependence on the molecular
length $L$, stick-slip persisting to the highest critical velocity for
$L=5$.
In the low-$v_s$ regime the friction data tend to become noisy, announcing
the well-known difficulty of MD to address the low shear rates of realistic
lab experiments.
For an extension of the present study to that regime, a transient-time
correlation function approach \cite{Delhommelle05} would be of substantial
help.

\begin{figure}
\includegraphics[width=82mm]{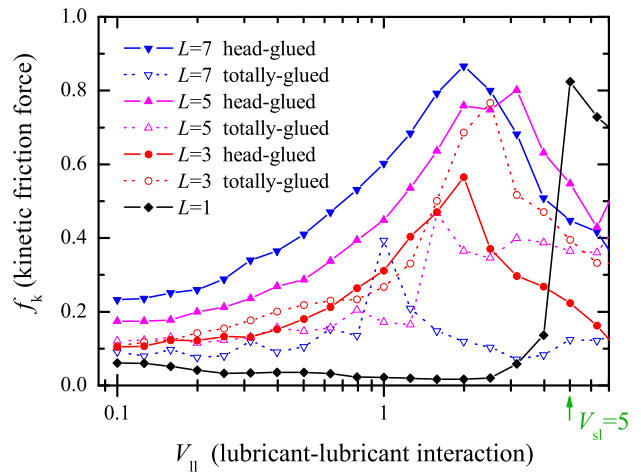}
\caption{\label{L11} (Color online)
Kinetic friction force $f_k$ as a function of the energy
amplitude $V_{ll}$ of interaction
between the lubricant atoms,
for linear lubricant molecules of different lengths
$L=1$ ($M=N=360$, solid diamonds),
$L=3$ ($M=120$ and $N=360$, red circles),
$L=5$ ($M=70$ and $N=350$, magenta up triangles), and
$L=7$ ($M=51$ and $N=357$, blue down triangles),
for either totally- (open symbols and dotted curves) or head-glued (solid
symbols and curves) cases.
Simulations carried out at temperature $T=0.1$ and driving velocity
$v_s=0.1$; other parameters same as in Fig.~\ref{L01}.}
\end{figure}

Figure~\ref{L11} demonstrates the dependence of the kinetic friction
on the strength of the interaction  $V_{ll}$ between the lubricant molecules.
At small $V_{ll}$, e.g., for $V_{ll}=0.1$, the surfaces are covered
by monolayers of lubricant molecules, and the sliding interface lies
somewhere near the middle of the lubricant film.
As $V_{ll}$ increases, sliding becomes more and more viscous,
thus friction increases reaching a maximum near
$V_{ll} = V_{ll}^* \alt V_{sl}=5$.
At this point, the sliding mechanism changes: the lubricant becomes rigid
enough to move as a whole body, and frictional sliding is now shifted to the
substrate/lubricant interfaces.
If $V_{ll}$ increases further beyond this point, sliding becomes easier and
easier because of increasing lubricant rigidity; this regime
corresponds to the sliding of two hard incommensurate surfaces \cite{P0,BN2006}.
We found the same type of crossover for other numbers of lubricant molecules.

\begin{figure}
\includegraphics[width=82mm]{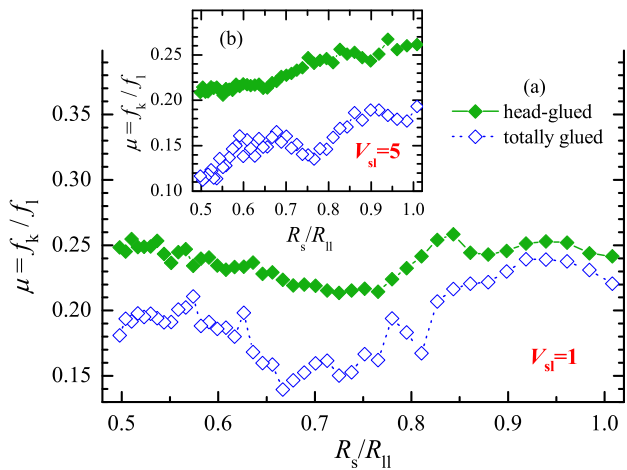}
\caption{\label{L12} (Color online)
Friction coefficient $\mu=f_k /f_l$ as a function of the substrate lattice
constant $R_s$ for $M=70$ linear lubricant molecules of length $L=5$ for
totally glued (blue open symbols) and head-glued (green solid symbols) chains.
The sliding velocity is $v_s=0.316$, the interaction radius $R_{sl}=R_s$ is
varied, and in (b) all other parameters are as in Fig.~\ref{L01} including
$V_{sl}=5$ or 1/9, while in (a) the gluing interaction
is $V_{sl}=1$ or 1/9.}
\end{figure}

Finally, Fig.~\ref{L12} demonstrates the variation of the friction
coefficient with the substrate lattice constant $R_s$.
In these simulations, we keep the size of the system constant at $L_x = 62
\times 2/3 \approx 41.3$, and vary the number of substrate atoms
from $N_s=41$ to~83, so that $R_s = L_x /N_s$ changes from $\approx 0.5$ to~$\approx 1$.
The total load force $F_l =f_l N_s$ is also kept constant at the level
$F_l =62$ (so that we have $f_l =1$ for $N_s =62$ as in the previous simulations).
As expected, there appears to be no systematic dependence of friction on
the substrate/lubricant interface $R_s/R_{ll}$ for the head-glued system,
where the molecules arrange themselves at frequently varying mutual
distance, and no commensurability effect can arise.

\begin{figure}
\includegraphics[width=82mm]{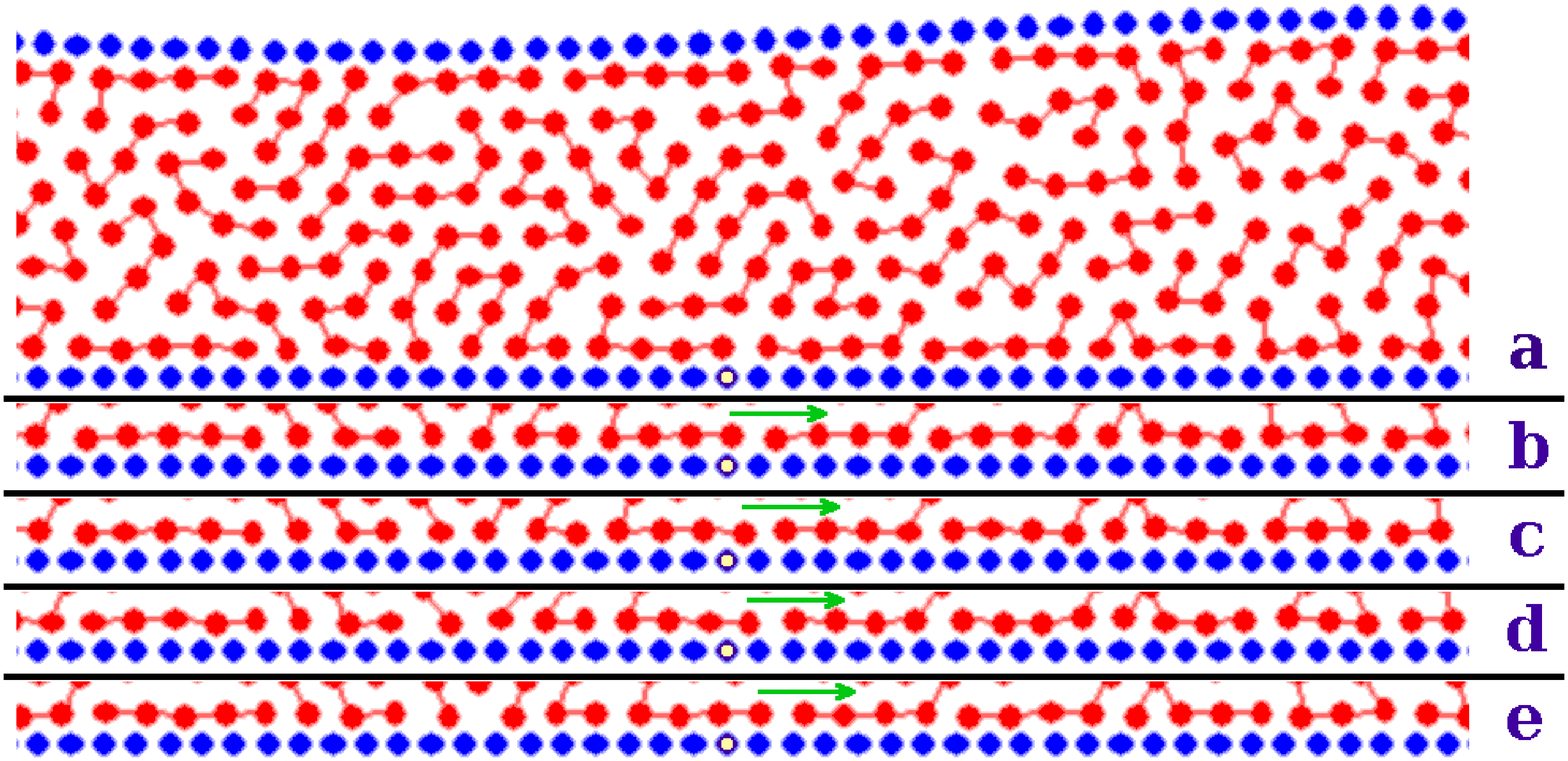}
\caption{\label{caterpillar} (Color online)
The central region of five successive snapshots of the motion of $M=70$
totally-glued $L=5$ lubricant molecules.
The time interval between successive snapshots (a) to (e) is $\Delta t=1$.
Snapshots b-e display only the atoms nearest to the bottom substrate,
for better visibility of the caterpillar motion.
To provide a clearer reference frame, one of the static atoms in the
bottom-substrate is drawn as empty.
The other parameters are as in Fig.~\ref{L12}(a), except for
$R_{sl}=R_s=2/3$.}
\end{figure}

For the totally glued case, however, the friction coefficient shows a well
pronounced minimum at $R_s \alt 0.75$, especially for a lower value of the
lubricant-substrate interaction $V_{sl} =1$.
This is associated to a slow ``caterpillar'' motion of the glued molecules
in the periodic substrate potential \cite{B1990}, caused by advancing of
kinks (misfit dislocations) at the lubricant-substrate interfaces
\cite{Vanossi06,Cesaratto07,Vanossi07PRL}.
For a rigid linear molecule in a sinusoidal potential, virtually free
sliding of the molecule should be observed for ratios $R_s/R_{\rm mol} =
L/(2L-n)$, where $n=1,2,\ldots,{\rm int} [L/2]$.
This gives, e.g., $R_s/R_{\rm mol}=2/3$, or $R_s \approx 0.57$ for $L=2$ and~4, and
$R_s/R_{\rm mol}=5/8$, or $R_s \approx 0.535$ for $L=5$.
For a nonrigid molecule, these values are
shifted upwards~\cite{B1990}.
As shown in Fig.~\ref{caterpillar},
a caterpillar-type motion indeed takes place.
However, as some molecules are attached with all $L$ atoms but others
stick only partially, e.g., with $2-4$~atoms, while the glued layer should
move as a whole (the molecules cannot overtake one another), caterpillar
motion is disturbed, and this makes the minimum in Fig.~\ref{L12} broad and shallow.
The dependence of friction on the incommensurability between the substrate
and molecular spacings was actually observed experimentally \cite{KM-05}.

\section{Discussion and conclusion}
\label{discussion}

We presented simulations of a simple 2D model of molecular lubrication.
Results indicate that layers of similar
thickness of totally-glued lubricant molecules generally produce a smaller
friction than lubricant molecules that are head-glued.
Totally-glued lubricants also stick better to the substrates, and
boundary layers thereof may accordingly be harder to squeeze out under high load.
On the other hand, whenever the applied load becomes strong enough to
squeeze the lubricant out to less than two molecular layers, then the
head-glued kind behaves much better in sustaining smooth sliding and low
friction even at very small residual coverage.
We predict that when such extreme-load situations are likely to occur in a
practical lubricated sliding devices, head-glued molecules could
provide the best
lubricant, unless the squeeze-out resistance is much worse than for totally
glued lubricant.
Moreover, when the base lubricant is totally-glued as a typical oil
and, therefore, provides low friction properties,
then even  a small concentration of
head-glued additives such as diblock polymers
may prevent the substrates from wearing at places (asperities)
where the surfaces come close to each other
within a thickness of less than two layers. 
This is a new result, susceptible of experimental test.

Similarly, friction shows a clear increasing trend as a function of
molecular length -- the number $L$ of monomer units in the molecules.
This result is similar to the $L$ dependency of the bulk viscosity, and
our results compare well to those of Ref.~\cite{SSP2003}.
This would make monoatomic liquids the best lubricants, if it was not for
their much reduced resistance against squeeze-out, which eventually makes
polyatomic chains more effective lubricants at loads where boundary
lubrication matters.

The detailed search for the best compromise between the necessity of a
squeeze-out resistance, calling for large $L$, and of low-friction
characteristics, calling for small $L$, is a rather intricate task that is
probably best undertaken on a case-by-case base by experimental trial and
error procedures, or possibly by very extended simulations of sophisticated
3D realistic models.
The simple 2D model at hand allows us to understand the main mechanisms at
play in the soft lubricant and to predict the general trends.
For example, local extra heating induced by the increased lubricant shear
velocity at thinnest lubricant regions between substrate corrugations may
well act as an important mechanisms producing a considerable overall friction
reduction in macroscopic sliding.

\acknowledgments
We wish to express our gratitude to
G.E.\ Santoro, A.\ Vanossi, I.E.\ Castelli, and R.\ Capozza
for useful discussions.
This research was supported in part by the
Central European Initiative (CEI), and by
a grant from the Cariplo Foundation
managed by the Landau Network -- Centro Volta,
whose contributions are gratefully acknowledged.
Work in SISSA sponsored by Italian Ministry of University and Research,
through PRIN-2006022847.



\begin{thebibliography}{11}

\bibitem{P0} B.N.J.\ Persson,
  {\it Sliding Friction: Physical Principles and Applications}
  (Springer-Verlag, Berlin, 1998).

\bibitem{BN2006} O.M.\ Braun and A.G.\ Naumovets, Surf.\ Sci.\ Reports {\bf
  60}, 79 (2006).

\bibitem{RT1991}  M.O.\ Robbins and P.A.\ Thompson, Science {\bf 253}, 916
  (1991).

\bibitem{SSP2003} I.M.\ Sivebaek, V.N.\ Samoilov, and B.N.J.\ Persson,
  J.\ Chem.\ Phys.\ {\bf 119}, 2314 (2003).

\bibitem{HR2001s} G.\ He and M.O.\ Robbins, \prb {\bf 64}, 35413 (2001).
\bibitem{persson_tosatti94} B.N.J.\ Persson and E.\ Tosatti, \prb {\bf 50},
  5590 (1994).

\bibitem{Gardiner}
C.W. Gardiner, \textit{Handbook of Stochastic Methods for Physics,
Chemistry and the Natural Sciences} (Springer-Verlag, Berlin, 1985).

\bibitem{MR2000} M.H.\ M\"user and M.O.\ Robbins, \prb {\bf 61}, 2335 (2000).
\bibitem{KG1990} K.\ Kremer and G.S.\ Grest, J.\ Chem.\ Phys.\ {\bf 92},
  5057 (1990).

\bibitem{T1998} W.\ Tsch\"op, K. Kremer, J. Batoulis, T. Burger, and O. Hahn,
  Acta Polym.\ {\bf 49}, 61 (1998);
  W.\ Tsch\"op, K. Kremer, O. Hahn, J. Batoulis, and T. Burger,
  Acta Polym.\ {\bf 49}, 75 (1998).
\bibitem{BP2003} O.M.\ Braun and M.\ Peyrard, \pre {\bf 68}, 011506 (2003).
\bibitem{Seeton2006} C.J.\ Seeton, Tribology Lett. {\bf 22}, 67 (2006).
\bibitem{Delhommelle05}J.\ Delhommelle and P.\ T.\ Cummings \prb {\bf 72},
  172201 (2005).
\bibitem{B1990} O.M.\ Braun, Surface Sci.\ {\bf 230}, 262 (1990).
\bibitem{Vanossi06} A.\ Vanossi, N.\ Manini, G.\ Divitini, G.E.\ Santoro,
  and E.\ Tosatti, \prl {\bf 97}, 056101 (2006).
\bibitem{Cesaratto07} M.\ Cesaratto, N.\ Manini, A.\ Vanossi, E.\ Tosatti,
  and G.E.\ Santoro, Surf.\ Sci.\ {\bf 601}, 3682 (2007).
\bibitem{Vanossi07PRL} A.\ Vanossi, N.\ Manini, F.\ Caruso, G.E.\ Santoro,
  and E.\ Tosatti, \prl {\bf 99}, 206101 (2007).
\bibitem{KM-05}
  V.S.\ Kulik, A.A.\ Marchenko, A.G.\ Naumovets, and J.\ Cousty,
  in: V.E.\ Borisenko, S.V.\ Gaponenko, and V.S.\ Gurin (eds.),
  Physics, {\it Chemistry and Application of Nanostructures}
  (World Scientific, Singapore, 2005) p.\ 74.
\end{thebibliography}
\end{document}